\documentclass[aps,a4paper,tightenlines,superscriptaddress]{revtex4}
\usepackage{amsmath,amssymb}
\usepackage{psfrag,verbatim}
\usepackage[dvips]{graphicx}
\usepackage{color,euler}

\begin{document}
Corresponding Author: Assistant Professor Sergey E. Strigin

Corresponding Author's Institution: Physics Department, Moscow
State University, Moscow, Russia 119991 GSP-1, Leninskie gory 1,
bld. 2, chair of physics of oscillations

Tel: +7(495)9394428

Fax: +7(495)9328820

E-mail address: strigin@phys.msu.ru
\newpage

%opening
\title{Analysis of parametric oscillatory instability in Fabry-Perot cavity with Gauss and Laguerre-Gauss main mode
profile}%of  gravitational wave Advanced VIRGO interferometer}
\author{S. E. Strigin and S. P. Vyatchanin}
\affiliation{Physics Department, Moscow State University, Moscow
119991 Russia}
%\date{ \today}

\begin{abstract}
We calculate the parametric instabilities in Fabry-Perot cavities
of Advanced VIRGO and LIGO interferometers with different main
mode profiles. All unstable combinations of elastic and Stokes
modes both for the case with $TEM_{00}$ and $LG_{33}$ as a
carriers are deduced.
\end{abstract}
\maketitle



\section{Introduction}

%The effect of parametric oscillatory instability is a dangerous one for gravitational wave detectors and we have to have any necessary information about how to avoid it.

%\section{Effect of parametric oscillatory instability}\label{PI}

Sensitivity increase of gravitational wave detectors like Advanced
VIRGO and LIGO is expected to be obtained by rising the value of
power  stored in the Fabry-Perot(FP) resonator optical mode.
However, high values of circulating power may be a source of the
nonlinear effects which will prevent from reaching the projected
sensitivity. It is appropriate to remind that nonlinear coupling
of elastic and light waves in continuous media produces
Mandelstam-Brillouin scattering. It is a classical parametric
effect; it is often explained in terms of quantum physics: one
quantum $\hbar \omega_0$ of main optical wave transforms into two,
i.e., $\hbar \omega_1$ in the additional optical wave (Stokes
wave: $\omega_1<\omega_0$) and $\hbar \omega_m$ in the elastic
wave so that $\omega_0=\omega_1+\omega_m$. The irradiation into
the anti-Stokes wave is also possible
($\omega_{1a}=\omega_0+\omega_m$), however, in this case the part
of energy is taken from the elastic wave. The physical mechanism
of this coupling is the dependence of refractive index on density
which is modulated by elastic waves. If the main wave power is
large enough the stimulated scattering will take place, the
amplitudes of elastic and Stokes waves will increase
substantially.

In gravitational wave detectors elastic oscillations in FP
resonator mirrors will interact with optical ones being coupled
parametrically due to the boundary conditions on the one hand, and
due to the ponderomotive force on the other hand. Two optical
modes may play roles of the main and Stokes waves. High quality
factors of these modes and of the elastic one will increase the
effectiveness of the interaction between them and may give birth
to the parametric oscillatory instability which is similar to
stimulated Mandelstam-Brillouin effect. This undesirable effect of
parametric oscillatory instability may create a specific upper
limit for the value of power $W_c$ circulating inside the
cavity\cite{bsv}.

It is interesting that the effect of parametric instability is
important not only for large scale gravitational-wave
interferometers. Recently, the instability produced by optical
rigidity was observed in experiment \cite{corbitt}.  K. Vahala
with collaborators has also observed it in micro scale whispering
gallery optical resonators \cite{kipp, rok}. Zhao et
al.\cite{zhao} have shown in experiment, using an 80 m Fabry-Perot
cavity, that parametric instability effect can indeed occur. The
experimental results are in good agreement with theoretical
predictions\cite{bsv}.

The condition of parametric instability for Fabry-Perot cavity may
be written in simple form\cite{bsv}:
\begin{eqnarray}\label{00}{{\cal R}=\frac{\Lambda_1 W
\omega_1}{cLm\omega_m\gamma_m\gamma_1}\times
\frac{1}{1+\frac{\Delta^2}{\gamma_1^2}}>1,\quad \gamma_1\gg
\gamma_m,}\\\label{o} {\Lambda_1=\frac{V(\int
A_0(r_{\bot})A_1(r_{\bot})u_z dr_{\bot})^2}{\int|A_0|^2
dr_{\bot}\int|A_1|^2dr_{\bot}\int|\vec{u}|^2 dV}},
\end{eqnarray}
where $c$ is a speed of light, $L$ is a distance between the FP
mirrors, $m$ is a mirror's mass, $\gamma_m,\,\gamma_1$ are the
relaxation rates of elastic and Stokes modes correspondingly, $W$
is a power circulating inside the cavity, $\Lambda_1$ is an
overlapping factor of elastic and optical modes and
$\Delta=\omega_0-\omega_1-\omega_m$ is a detuning value. Here
$A_0$ and $A_1$ are the functions of the distributions over the
mirror surface of the optical fields in the main and Stokes
optical modes correspondingly, vector $\vec{u}$ is the spatial
vector of displacements in elastic mode, $u_z$ is the component of
$\vec{u}$, normal to the mirror's surface, $\int d{r}_{\bot}$
corresponds to the integration over the mirror surface and $\int
dV$ --  over the mirror volume $V$.

As we can see from formula (\ref{00}) the parametric instability
is a threshold effect and it takes place if optical power $W$ in
main mode of Fabry-Perot cavity is bigger than the threshold power
$W_c$. The value $W_c$ depends on the detuning $\Delta$ between
optical and elastic modes and overlapping factor $\Lambda_1$, in
particular, the $W_c$ has minimum if detuning  is less than
relaxation rate of  Stokes mode  $\Delta\ll \gamma_1$ and
overlapping factor is large enough.
\begin{figure}[h]
%\begin{picture}
%\begin{center}
%\epsfxsize=3.2in \epsfbox{compens.eps}
\psfrag{w0}{$\omega_0$} \psfrag{w1}{$\omega_1$} \psfrag{?}{?}
\psfrag{main}{Main modes}
\includegraphics[width=9cm,totalheight=4cm]{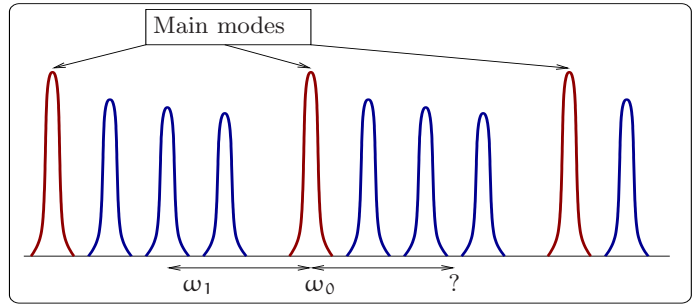}
\caption{ Schematic structure of optical Laguerre-Gauss modes in
the FP cavity. The modes of the main frequencies sequence are
shown by higher peaks. It is shown that Stokes mode with frequency
$\omega_1$ may not have suitable anti-Stokes mode (it is denoted
by question-mark).}\label{modes}
%\end{center}
%\begin{picture}
\end{figure}

Analyzing this  equation D'Ambrosio and Kells \cite{ak} pointed
out  that the presence of the anti-Stokes mode can considerably
depress or even exclude parametric instability. Indeed, the
condition of parametric oscillatory instability in the presence of
anti-Stokes mode with frequency $\omega_{1a}=\omega_0+\omega_m$
has the following form\cite{bsv2}: $${\frac{\Lambda_1 W
\omega_1}{cLm\omega_m\gamma_m\gamma_1}\times\frac{1}{{1+\frac{\Delta^2}{\gamma_1^2}}}-\frac{\Lambda_{1a}
W\omega_1}{cLm\omega_m\gamma_m\gamma_1}\times\frac{\omega_{1a}\gamma_1}{\omega_1\gamma_{1a}}\times\frac{1}{{1+\frac{\Delta_{1a}^2}{\gamma_{1a}^2}}}>1,}$$
where $\gamma_{1a}$ is a relaxation rate of anti-Stokes mode and
$\Delta_{1a}=\omega_{1a}-\omega_0-\omega_m$ is a detuning value.

 For example, let the main, Stokes and anti-Stokes modes be equidistant and belong to the main frequency
 sequence $\omega_1=\pi(K-1)c/L$, $\omega_0=\pi Kc/L$, $\omega_{1a}=\pi(K+1)c/L$(K is an integer). In this
 case $\Delta=\Delta_{1a}$, the main, Stokes and anti-Stokes modes have the same Gaussian
distribution over the cross section and hence the same overlapping
factors: $\Lambda_1 = \Lambda_{1a}$. It means that the second term
in the right part of equation is larger than first term, the
positive damping introduced into elastic mode by the anti-Stokes
mode is greater than negative damping due to the Stokes mode,
hence the parametric instability is impossible. This case has been
analyzed in details in \cite{ak}.

For the case when the Stokes and anti-Stokes modes do not belong
to the main sequence (non-zero radial $N$ and  azimuth $M$
numbers) the frequencies of the suitable Stokes and anti-Stokes
modes are not equidistant from the main mode ($\Delta \neq
\Delta_{1a}$) and have different spatial distributions ($\Lambda_1
\neq \Lambda_{1a}$).  It takes place when main mode is $TEM_{00}$
and Fig. \ref{modes} illustrates it. For the shown Stokes mode
(left to the main mode) there is no suitable anti-Stokes mode (it
should be located right to the main one). The probability that
appropriate anti-Stokes mode exists is extremely small and we see
that in this case full depression of parametric instability does
not take place.  However, when main mode has non-zero radial and
azimuth numbers(high order Laguerre-Gauss modes $LG_{pn}$) the
possibility of parametric instability depression by existence of
suitable anti-Stokes mode will increase.

Parametric instability is a serious problem for advanced
gravitational wave detectors and we need to know this ''enemy'' in
detail to avoid it. In particular, we need detailed information
about pairs of Stokes modes and elastic modes which may be
possible candidates for parametric instability. It is known that
Stokes and anti-Stokes modes may be analytically calculated for
Gaussian beams.  In contrast, elastic modes may be calculated
numerically(some method of analytical elastic modes calculation
has been proposed in \cite{meleshko}). Hence, the accuracy of
parametric instability forecast (and success of methods to prevent
it) directly depends on how accurately we can calculate normal
(eigen) frequencies and spatial distributions of elastic modes.
Using ANSYS$^\circledR$ code, for example, we can obtain the
accuracy of elastic modes calculations of about 0.5$\%$ or
poorer\cite{1b,2b,3b,4b}.

In \cite{striginblair,strigin}  the importance of accurate
numerical calculation of elastic modes in the mirrors of Advanced
LIGO was discussed to enable precise predictions of the problem of
parametric oscillatory instability. In \cite{striginblair} it has
been proposed accuracy estimations through use of analytical
solutions based on Chree-Lamb modes. Small deviations from
cylindrical test mass shape may produce splitting of non-axial
symmetric elastic modes into doublets. This splitting may increase
the possibility of parametric oscillatory instability
too\cite{strigin}.

 It is known the important goal in gravitational wave detectors is to minimize fundamental and technical noise contribution.
 For example, progressive way to lower the thermal noise is to change the mode shape of the laser beam inside the
 interferometer. It is worth noting that different shapes of beam have been proposed for reducing thermal noise such as
 mesa beams \cite{thorne,thorne1,bond}, conical modes \cite{bond1} and high order Laguerre-Gauss modes\cite{mours}.
%In section \ref{PI} we  recall the physical sence and appropriate formulas  of this effect for FP cavity which permit to obtain approximate estimates for the instability conditions.%
At present, there are some techniques for the generation of high
order LG modes using holograms\cite{arlt,clif},
gratings\cite{kenn} and mode transformers\cite{cour,neil} with
high conversion efficiency\cite{kenn,chu}. In \cite{hild} using
numerical interferometer simulations the comparison of behaviour
of the $LG_{33}$ mode with the fundamental mode $TEM_{00}$ are
discussed. The $LG_{33}$ mode performs similar if not even better
than commonly used $TEM_{00}$ for all considered aspects of
interferometric sensing.

 In this Letter
we  estimate the possibility of parametric instabilities in
Fabry-Perot cavity of Advanced VIRGO and LIGO interferometers both
for Gauss $TEM_{00}$ and Laguerre-Gauss $LG_{33}$ modes as a
carriers\cite{hild}. In section \ref{results} we  provide
estimations of PI in Fabry-Perot cavity of Advanced VIRGO and LIGO
interferometers and discuss our results, in section
\ref{conclusion} we make some useful conclusions for future
research.

\section{Results}\label{results}

In order to predict the unstable combinations of Stokes and
elastic modes we have to take into account the additional azimuth
numbers condition of parametric instability. For example, for
$TEM_{00}$ as a carrier (when the elastic modes have azimuth
dependence $\sim e^{im\phi}$ and Stokes modes have - $e^{in\phi}$)
the non-zero overlap factor will be if $m = n$(see the formula
(\ref{o}) for  $\Lambda_1$). In turn, if we use $LG_{33}$ as a
carrier  the non-zero overlap factor will be if  $|m\pm n|=3$. It
is worth noting that it is correct only if cylinder center
coincides with laser spot center and below we consider this
particular case(in opposite case the overlap factor has to depend
on distance between the center of mirror and center of main
optical mode distribution over mirror surface).

 For Fabry-Perot cavity we use the parameters of Advanced VIRGO and LIGO interferometers shown in Table \ref{table4}
\begin{table}[h]
\caption{Parameters of Advanced LIGO and VIRGO interferometers}
\label{table4} \vskip15pt
\begin{center}
%\footnotesize{
\begin{tabular}{|c|c|c|c|c|c|}
%\hline
%& \multicolumn{2}{c|}{?" ?"?e� ?? a?}
%&\multicolumn{2}{c|}{' ?a?a}\\
\hline   VIRGO & LIGO
\\ \hline \hline%$2 \pi\times$ (5831.6$\pm$)& $2 \pi\times$ (5855.8$\pm$) & $ 2$ \\
  %$2 \pi\times$(5945.9$\pm$) &  $2 \pi\times$ (5855.2$\pm$)  &  $2$ \\ \hline
 $W=0.76\times 10^6 W$  & $W=0.83\times 10^6 W$
\\
$L=3000m$ &$L=4000m$  \\ $T=7\times 10^{-3}$ &$T=14\times 10^{-3}$
\\ $\lambda=1064nm$ &$\lambda=1064nm$  \\ $m=40kg$ & $m=40kg$ \\
$r=0.17m$ &$r=0.17m$\\ $H=0.2m$& $H=0.2m$\\ $\gamma_m=6\times
10^{-3}s^{-1}$ & $\gamma_m=6\times 10^{-3}s^{-1}$\\
$\gamma=cT/4L=175s^{-1}$ & $\gamma=cT/4L=262.5s^{-1}$\\
%\hline 25075 & 0 & $LG_{53}$& 108.9  \\
%\hline 25075 & 0 & $LG_{53}$& 108.9  \\
%\hline
 %  &$5.5440$ & $2 \pi\times$(19539.07$\pm$0.04) & $5.5422$
%\\\hline
%$2 \pi\times$ (24915$\pm$20)& $7.0672$ & $2 \pi\times$  (24896.78$\pm$0.04)&
%$7.0620$\\
%7436.7763   &  7442.4390 & 8011.8079 & -&
% $\varphi$  & 7758.6870  &  7780.9456 & 9220.0149 & -&
%$\varphi$  & 7966.4337  &  7807.2823 & 9220.0149 & -& ?4$\varphi$
%& 8133.8298 & 8218.8056 & 8257.11 & -&
% 3$\varphi$ & 8308.3169 &
%8470.4282 & - & -& $\varphi$ & 8503.0527 & 8567.8158 & 10068.9966
%& -&
% 3$\varphi$ & 8505.6099  &  8534.1361 & - &-&
% ?$\varphi$  & 8549.1754  &  8585.5572 & 9913.5688 & -&
% axial-symmetric  & 9267.6096  &  9225.5442 & 10500.8450 & -&
% 3$\varphi$ & 10409.8279  &  10444.7837 & 10144.0310 & -&
%3$\varphi$ & 10481.0205  &  10489.7286 & 10153.5094 & -&
%2$\varphi$& 10804.6306  &  10823.0441 & - & -& 2$\varphi$ &
%10991.9418  & 11070.2173 & - & -&
% $\varphi$ & 11114.9279 &
%11230.8881 & - & -& $\varphi$ & 11222.2805  &  11370.8341 & - & -&
% 2$\varphi$ & 11304.1022  &  11452.8592 & - & -&
% 2$\varphi$  & 11371.1159  &  11494.2472 & - & -&
% 4$\varphi$  & 11602.5558 &  11962.8212 & - & -& 4$\varphi$ &
%11859.1398  & 12060.1185 & - & -&
\hline
\end{tabular}%}
\end{center}
\end{table}
%\begin{align*}
% W_{VIRGO} = 0.76\times 10^6 {\rm W},\,W_{LIGO} = 0.83\times 10^6 {\rm W},\, L_{VIRGO}=3000{\rm m},\,L_{LIGO}=4000{\rm %m},\\ T_{VIRGO}=7\times 10^{-3},\,T_{LIGO}=5\times 10^{-3},\lambda=1064{\rm nm},\,\\
 %m \simeq 40{\rm kg},\,r=0.17{\rm m},\, H=0.2{\rm m},\,\gamma_m=6\times 10^{-3}{\rm s}^{-1},\gamma=cT/4L,
%\end{align*}
where $T$ are the transmittances  of input FP mirrors, $\lambda$
is a wave length of input laser and $r, H$ are radius and height
of cylindrical fused silica mirror. Note that the condition of
parametric instability in Michelson interferometer with
Fabry-Perot cavities in the arms with additional power and signal
recycling mirrors(configurations of Advanced LIGO and VIRGO
interferometers) and in alone Fabry-Perot cavity can substantially
differ -- see discussion in Conclusions.

The calculation of parametric instabilities is very sensitive to
small changes in the model parameters and gives only statistical
result for number of unstable modes. In particular, uncertainties
in the mirror radius of curvature and small variations in
materials and temperature are sufficient to move the optical and
mechanical resonances of the cavities.
 We  estimate the number of unstable combinations of elastic and Stokes modes  both for the case  with $TEM_{00}$ as a
 carrier($TEM_{00}$ case) and --  $LG_{33}$ as a carrier($LG_{33}$ case). The elastic modes were calculated numerically
 using COMSOL$^\circledR$ code on triangle mesh with about 40000 meshing elements. We restrict our estimates by elastic
 frequency range up to $2\pi\times 40000 s^{-1}$ because, on the one hand, the more the elastic frequency the less the
 parametric gain. On the other hand, high elastic  mode distributions are difficult to determine exactly among other
 elastic modes due to degradation of numerical accuracy of COMSOL$^\circledR$ code.

 For VIRGO configuration we use Fabry-Perot cavities with laser spot radii on mirrors surfaces
 $w = 6.47$~cm and $w=3.94$~cm in $TEM_{00}$ and $LG_{33}$ cases correspondingly\cite{hild}. The
 waist radii are situated in the middle of the cavities and are equal to $w_0\simeq 0.79 $cm and $w_0\simeq 1.38$cm for
 these cases. Level of diffractional losses in clipping approximation is hold to be $l_{clip} = 1$~ppm. For LIGO
 configuration we use Fabry-Perot cavities with laser spot radii on mirrors surfaces $w = 6$~cm and $w=3.94$~cm in
 $TEM_{00}$ and $LG_{33}$ cases correspondingly\cite{abbott}. The waist radii are situated in the middle of the cavities
 and are equal to $w_0\simeq 1.15$cm and $w_0\simeq 1.99$cm for these cases, and level of diffractional losses in clipping
 approximation is hold to be $l_{clip} = 1$~ppm or less.

 We calculate the values of overlap factors for all combinations of elastic and optical modes(up to $9^{th}$ order) but,
 on the other hand, we use azimuth numbers condition in each case. It is also worth noting that the analysis of
 parametric instability in each case has been realized with  taking into account the appropriate anti-Stokes modes
 which may suppress the parametric instabilities.  We also took into account the diffractional losses of Stokes and
 anti-Stokes modes in clipping approximation for estimations of optical modes relaxation rates as it has been done
 in \cite{bsv2}.

\begin{figure}
%\begin{picture}
\begin{center}
\includegraphics[width=7cm,totalheight=4.7cm]{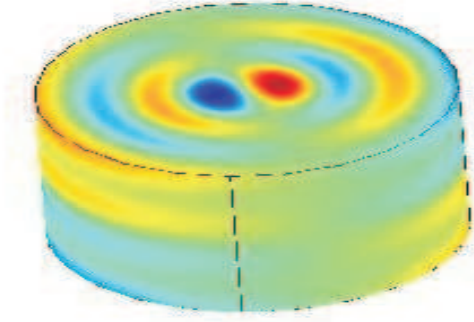}
\caption{Displacement vector component distribution $u_z$ of
elastic mode with frequency $38577{\rm Hz}$  and azimuth index
$m=1$. The numerical calculations of cylinder  have been made on
triangle mesh with about 40000 meshing elements. }\label{fig1}
\end{center}
%\begin{picture}
\end{figure}

In Table \ref{table1} all  unstable combinations of elastic and
Stokes modes and parametric gains $R$ for $LG_{33}$ case in VIRGO
interferometer are shown. There is only one unstable combination
and there is not any anti-Stokes mode that can suppress the
parametric instability. In figure \ref{fig1} we can see the
z-component displacement vector distribution for this unstable
elastic mode. In the case of $TEM_{00}$ carrier we did not find
any  unstable modes in given frequency range due to small overlap
factors and large detuning values $\Delta$.

\begin{figure}[h]
%\begin{picture}
\begin{center}
\includegraphics[width=7cm,totalheight=5cm]{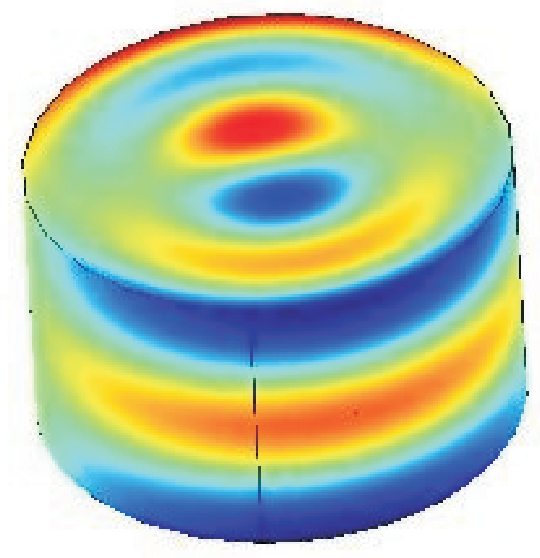}
\includegraphics[width=7cm,totalheight=5cm]{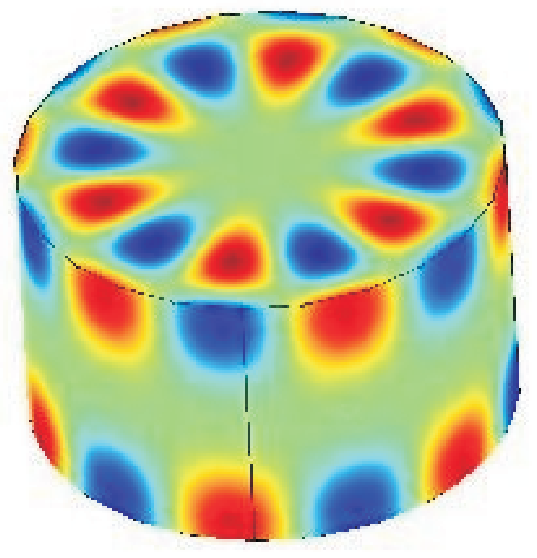}
\includegraphics[width=7cm,totalheight=5cm]{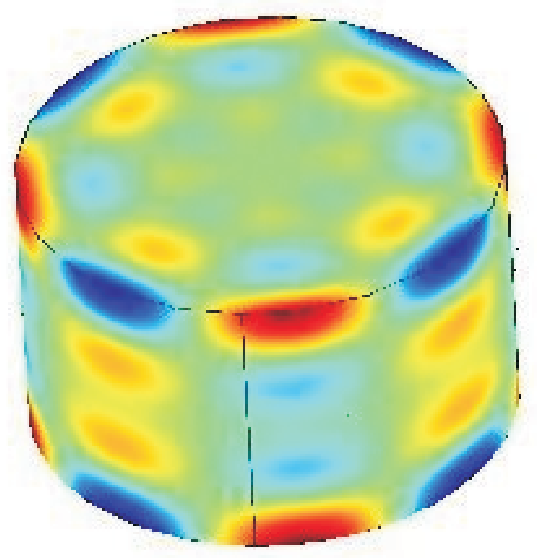}
\caption{Displacement vector component distributions $u_z$  of
elastic modes with frequencies $23048{\rm Hz}$(azimuth index
$m=1$), $36524{\rm Hz}$(azimuth index $m=6$) and $37566{\rm
Hz}$(azimuth index $m=4$) correspondingly. The numerical
calculations of cylinder  have been made on triangle mesh with
about 40000 meshing elements.}\label{fig2}
\end{center}
%\begin{picture}
\end{figure}

In Table \ref{table2} all  unstable combinations of elastic and
Stokes modes and parametric gains $R$ for $LG_{33}$ case in LIGO
interferometer are shown. There are three  unstable combinations.
In figure \ref{fig2} we present the z-component displacement
vector distributions for these  elastic modes.  In the case of
$TEM_{00}$ carrier we found only one unstable mode in given
frequency range which is shown in  Table \ref{table3}.

We can see that possibility of parametric instability in given
elastic frequency range for $LG_{33}$ case will be slightly larger
than for the $TEM_{00}$ case both in Fabry-Perot cavities of VIRGO
and LIGO interferometers.

\begin{table}[h]
\caption{Unstable combination of elastic and Stokes optical modes
in FP cavity of VIRGO configuration with $LG_{33}$ as a carrier.
In the case of $TEM_{00}$ carrier we did not find any  unstable
modes in given frequency range due to small overlap factors and
large detuning values $\Delta$.} \label{table1} \vskip15pt
\begin{center}
%\footnotesize{
\begin{tabular}{|c|c|c|c|c|c|}
%\hline
%& \multicolumn{2}{c|}{?" ?"?e� ?? a?}
%&\multicolumn{2}{c|}{' ?a?a}\\
\hline   $\omega_m/2\pi, {\rm Hz}$ & $m$    & Stokes optical mode
& $R$
\\ \hline %$2 \pi\times$ (5831.6$\pm$)& $2 \pi\times$ (5855.8$\pm$) & $ 2$ \\
  %$2 \pi\times$(5945.9$\pm$) &  $2 \pi\times$ (5855.2$\pm$)  &  $2$ \\ \hline
 38577  & 1     & $LG_{32}$ & 2.5
\\% \hline
%34360 &2
%& $LG_{12}$ & 8.2  \\
%\hline
%31264 &4
%& $LG_{24}$ & 2.6  \\
%\hline %$2 \pi\times$(15147$\pm$ 1) & $4.29650$ & $2 \pi\times$ (15147.015$\pm$ 0.004)& $4.29647$  \\
%\hline
 %   $2 \pi\times$(19545$\pm$ 6)&$5.5440$ & $2 \pi\times$(19539.07$\pm$0.04) & $5.5422$
%\\\hline
%$2 \pi\times$ (24915$\pm$20)& $7.0672$ & $2 \pi\times$  (24896.78$\pm$0.04)&
%$7.0620$\\
%7436.7763   &  7442.4390 & 8011.8079 & -&
% $\varphi$  & 7758.6870  &  7780.9456 & 9220.0149 & -&
%$\varphi$  & 7966.4337  &  7807.2823 & 9220.0149 & -& ?4$\varphi$
%& 8133.8298 & 8218.8056 & 8257.11 & -&
% 3$\varphi$ & 8308.3169 &
%8470.4282 & - & -& $\varphi$ & 8503.0527 & 8567.8158 & 10068.9966
%& -&
% 3$\varphi$ & 8505.6099  &  8534.1361 & - &-&
% ?$\varphi$  & 8549.1754  &  8585.5572 & 9913.5688 & -&
% axial-symmetric  & 9267.6096  &  9225.5442 & 10500.8450 & -&
% 3$\varphi$ & 10409.8279  &  10444.7837 & 10144.0310 & -&
%3$\varphi$ & 10481.0205  &  10489.7286 & 10153.5094 & -&
%2$\varphi$& 10804.6306  &  10823.0441 & - & -& 2$\varphi$ &
%10991.9418  & 11070.2173 & - & -&
% $\varphi$ & 11114.9279 &
%11230.8881 & - & -& $\varphi$ & 11222.2805  &  11370.8341 & - & -&
% 2$\varphi$ & 11304.1022  &  11452.8592 & - & -&
% 2$\varphi$  & 11371.1159  &  11494.2472 & - & -&
% 4$\varphi$  & 11602.5558 &  11962.8212 & - & -& 4$\varphi$ &
%11859.1398  & 12060.1185 & - & -&
\hline
\end{tabular}%}
\end{center}
\end{table}

\begin{table}[h]
\caption{Unstable combinations of elastic and Stokes optical modes
in FP cavity of LIGO configuration with $LG_{33}$ as a carrier. }
\label{table2} \vskip15pt
\begin{center}
%\footnotesize{
\begin{tabular}{|c|c|c|c|c|c|}
%\hline
%& \multicolumn{2}{c|}{?" ?"?e� ?? a?}
%&\multicolumn{2}{c|}{' ?a?a}\\
\hline   $\omega_m/2\pi, {\rm Hz}$ & $m$    & Stokes optical mode
& $R$
\\ \hline %$2 \pi\times$ (5831.6$\pm$)& $2 \pi\times$ (5855.8$\pm$) & $ 2$ \\
  %$2 \pi\times$(5945.9$\pm$) &  $2 \pi\times$ (5855.2$\pm$)  &  $2$ \\ \hline
 36524  & 6     & $LG_{03}$ & 1.5
\\ \hline
37566 &4 & $LG_{17}$ & 9.3  \\\hline 37566 &4 & $LG_{41}$ & 10.8
\\ %\hline 23048$^*$ &1 & $LG_{21}$ & 1.7  \\
%\hline 25075 & 0 & $LG_{53}$& 108.9  \\
%\hline 25075 & 0 & $LG_{53}$& 108.9  \\
%\hline
 %  &$5.5440$ & $2 \pi\times$(19539.07$\pm$0.04) & $5.5422$
%\\\hline
%$2 \pi\times$ (24915$\pm$20)& $7.0672$ & $2 \pi\times$  (24896.78$\pm$0.04)&
%$7.0620$\\
%7436.7763   &  7442.4390 & 8011.8079 & -&
% $\varphi$  & 7758.6870  &  7780.9456 & 9220.0149 & -&
%$\varphi$  & 7966.4337  &  7807.2823 & 9220.0149 & -& ?4$\varphi$
%& 8133.8298 & 8218.8056 & 8257.11 & -&
% 3$\varphi$ & 8308.3169 &
%8470.4282 & - & -& $\varphi$ & 8503.0527 & 8567.8158 & 10068.9966
%& -&
% 3$\varphi$ & 8505.6099  &  8534.1361 & - &-&
% ?$\varphi$  & 8549.1754  &  8585.5572 & 9913.5688 & -&
% axial-symmetric  & 9267.6096  &  9225.5442 & 10500.8450 & -&
% 3$\varphi$ & 10409.8279  &  10444.7837 & 10144.0310 & -&
%3$\varphi$ & 10481.0205  &  10489.7286 & 10153.5094 & -&
%2$\varphi$& 10804.6306  &  10823.0441 & - & -& 2$\varphi$ &
%10991.9418  & 11070.2173 & - & -&
% $\varphi$ & 11114.9279 &
%11230.8881 & - & -& $\varphi$ & 11222.2805  &  11370.8341 & - & -&
% 2$\varphi$ & 11304.1022  &  11452.8592 & - & -&
% 2$\varphi$  & 11371.1159  &  11494.2472 & - & -&
% 4$\varphi$  & 11602.5558 &  11962.8212 & - & -& 4$\varphi$ &
%11859.1398  & 12060.1185 & - & -&
\hline
\end{tabular}%}
\end{center}
\end{table}

\begin{table}[h]
\caption{Unstable combination of elastic and Stokes optical modes
in FP cavity of LIGO configuration with $TEM_{00}$ as a carrier. }
\label{table3} \vskip15pt
\begin{center}
%\footnotesize{
\begin{tabular}{|c|c|c|c|c|c|}
%\hline
%& \multicolumn{2}{c|}{?" ?"?e� ?? a?}
%&\multicolumn{2}{c|}{' ?a?a}\\
\hline   $\omega_m/2\pi, {\rm Hz}$ & $m$    & Stokes optical mode
& $R$
\\ %\hline %$2 \pi\times$ (5831.6$\pm$)& $2 \pi\times$ (5855.8$\pm$) & $ 2$ \\
  %$2 \pi\times$(5945.9$\pm$) &  $2 \pi\times$ (5855.2$\pm$)  &  $2$ \\ \hline
% 36524  & 6     & $LG_{03}$ & 1.5
%\\ \hline
%37566 &4 & $LG_{17}$ & 9.3  \\\hline 37566 &4 & $LG_{41}$ & 10.8
%\\\hline
\hline 23048 &1 & $LG_{21}$ & 1.7  \\
%\hline 25075 & 0 & $LG_{53}$& 108.9  \\
%\hline 25075 & 0 & $LG_{53}$& 108.9  \\
%\hline
 %  &$5.5440$ & $2 \pi\times$(19539.07$\pm$0.04) & $5.5422$
%\\\hline
%$2 \pi\times$ (24915$\pm$20)& $7.0672$ & $2 \pi\times$  (24896.78$\pm$0.04)&
%$7.0620$\\
%7436.7763   &  7442.4390 & 8011.8079 & -&
% $\varphi$  & 7758.6870  &  7780.9456 & 9220.0149 & -&
%$\varphi$  & 7966.4337  &  7807.2823 & 9220.0149 & -& ?4$\varphi$
%& 8133.8298 & 8218.8056 & 8257.11 & -&
% 3$\varphi$ & 8308.3169 &
%8470.4282 & - & -& $\varphi$ & 8503.0527 & 8567.8158 & 10068.9966
%& -&
% 3$\varphi$ & 8505.6099  &  8534.1361 & - &-&
% ?$\varphi$  & 8549.1754  &  8585.5572 & 9913.5688 & -&
% axial-symmetric  & 9267.6096  &  9225.5442 & 10500.8450 & -&
% 3$\varphi$ & 10409.8279  &  10444.7837 & 10144.0310 & -&
%3$\varphi$ & 10481.0205  &  10489.7286 & 10153.5094 & -&
%2$\varphi$& 10804.6306  &  10823.0441 & - & -& 2$\varphi$ &
%10991.9418  & 11070.2173 & - & -&
% $\varphi$ & 11114.9279 &
%11230.8881 & - & -& $\varphi$ & 11222.2805  &  11370.8341 & - & -&
% 2$\varphi$ & 11304.1022  &  11452.8592 & - & -&
% 2$\varphi$  & 11371.1159  &  11494.2472 & - & -&
% 4$\varphi$  & 11602.5558 &  11962.8212 & - & -& 4$\varphi$ &
%11859.1398  & 12060.1185 & - & -&
\hline
\end{tabular}%}
\end{center}
\end{table}

\section{Conclusions}\label{conclusion}
In this Letter we deduced the analysis of parametric oscillatory instability in FP cavities of  gravitational wave detectors like Advanced LIGO and VIRGO in the model with cylindrical mirrors
without flats and suspension ears. The main goal of our estimations was to compare a difference in $TEM_{00}/LG_{33}$ behavior between Advanced LIGO and VIRGO Fabry-Perot cavities for parametric instabilities in such simple model.  The number of obtained unstable modes for $LG_{33}$ case is slightly larger than for $TEM_{00}$ case  in elastic modes frequency range up to 40kHz, however, this difference is not large enough to conclude that $LG_{33}$ case is more dangerous than $TEM_{00}$ case.
No doubts, there is necessity to perform detailed analysis for full scale
schemes of Advanced LIGO and VIRGO in $LG_{33}$ case with wide
elastic modes frequency range, very high accuracy of elastic modes
calculations\cite{meleshko} and mirror flats and ears.

 At the end of the Letter we would like to summarize some dangerous points
that must be taken into  account at present and in future
parametric instability research.

 It is worth noting that  for combinations of modes suitable for parametric instability the overlapping
 factor $\Lambda_1$ may be zero (for example, elastic mode and the Stokes mode can have different dependence
 on azimuth angle)  --- we did not mention such combinations in our estimations. However, it is important to take
 into account that only the elastic mode is attached to the mirror axis in contrast to the optical mode which can be
 shifted from the mirror axis due to non-perfect optical alignment. Hence, the overlapping factor has to depend on
 distance $Z$ between the center of mirror and the center of the main optical mode distribution over the mirror surface.
 It means that $\Lambda_1$ may be zero for $Z = 0$ but non-zero for $Z\neq 0$. Therefore, the numerical analysis of the
 mode structure should evidently include the case when $Z\neq 0$. Note that there is a proposal to use special shift $Z$
 of the laser beam of about several centimeters from the mirror axis in order to decrease thermal suspension
 noise \cite{levin}.
%\item If cylinder center does not coincide with laser spot center (but $R\geq 1$ and $|m\pm n|= 3$) the overlap factor can be non-zero.

 Imperfections from the cylinder shape (such as flats and suspension ears), should cause splitting of elastic modes
 with azimuth index $m\geq 1$ into doublets and the difference between
doublet frequencies may be large enough or greater as compared
with relaxation rate of Stokes mode. Both the appearance of
doublets and high density of elastic modes may increase the
possibility of parametric instability in Advanced VIRGO and
LIGO\cite{striginblair,strigin}.

 The effect of parametric instability for power recycled
interferometer may be larger than for the separate Fabry-Perot
cavity because the Stokes mode emitted from the Fabry-Perot cavity
throughout its input mirror is not lost irreversible but returns
back due to power recycling mirror, therefore, its interaction is
prolonged\cite{bsv2}. As it was demonstrated in \cite{gbz} the
presence of power recycling cavity causes parametric gains $R$ to
be hugely amplified at $\Delta=0$ and reduced by a factor of 2 in
off-resonance intervals as compared to an interferometer without
power recycling. The detailed analysis of parametric instability
in signal recycled interferometer has been realized in
\cite{GSV,sv,sv1}. On the other hand, the parametric instability
in the interferometer with power recycling mirror may be depressed
by variation of distance between power recycling mirror and beam
splitter\cite{GSV,sv,sv1,cqg}. Detail analysis of parametric
instability in full scale Advanced LIGO interferometer has been
performed in\cite{cqg,frit} and it will be useful to present the
same analysis for $LG_{33}$ case in future researches.

We think that parametric oscillatory instability effect can be
overcome in laser gravitational detectors after developing
strategies for its suppression using these investigations.

\section*{ACKNOWLEDGEMENTS}

Authors are grateful to S. Hild, A. Freise and especially to M.L.~Gorodetsky for fruitful and stimulating discussions. 
Many thanks to R.~Frey and P.~Fritschel for their valuable remarks. S. E. Strigin
was supported by grant of President of Russian Federation No.
MK-195.2007.2, by grant of Moscow State University in 2008 and
2009 and by grant 08-02-00580-a of Russian Foundation for Basic
Research. S. P. Vyatchanin was supported by NSF grant PHY-0651036.


\begin{thebibliography}{99}
\bibitem{bsv}    V.\ B.\ Braginsky, S.\ E.\ Strigin, and S.\ P.\ Vyatchanin,
    {\em Physics Letters} {\bf A287}, 331 (2001);

\bibitem{corbitt} T. Corbitt, D. Ottaway, E. Innerhofer, J. Pelc, and N. Mavalvala,  Phys. Rev. A 74,
021802, 2006;
\bibitem{kipp} T. J. Kippenberg, H. Rokhsari, T. Carmon, A. Scherer, and K. J. Vahala, Phys. Rev.
Lett. 95, 033901, 2005;
\bibitem{rok} H. Rokhsari, T. J. Kippenberg, T. Carmon, K. J. Vahala, Optics express 13,
5293, 2005;
\bibitem{zhao} C. Zhao et al, Physical Review A 78, 023807, 2008;

   \bibitem {ak}
    E. D'Ambrosio and W. Kells,
    {\em Physics Letter } {\bf A299}, 326 (2002).
    \bibitem{bsv2}
V.\ B.\ Braginsky, S.\ E.\ Strigin and S.\ P.\ Vyatchanin,
    {\em Physics Letters} {\bf A305}, 111 (2002).

\bibitem{meleshko} V.V. Meleshko, S.E. Strigin, and M.S. Yakymenko, Physics Letters A373, 3701, 2009;
\bibitem{1b}C. Zhao, L. Ju, J. Degallaix, S. Gras and D.G. Blair, Phys. Rev. Lett 94, 121102, 2005;
\bibitem{2b}L. Ju, C. Zhao, S. Gras, J. Degallaix, D.G. Blair, J. Munch and D.H. Reitze, Phys. Lett. A 355, 419, 2006;
\bibitem{3b}L. Ju, S. Gras, C. Zhao, J. Degallaix,  and D.G. Blair, Phys. Lett A 354, 360, 2006;
\bibitem{4b}S. Gras, D.G. Blair, and C. Zhao, Classical and Quantum Gravity 26, 135012, 2009;

\bibitem{striginblair} S.E.Strigin, D.G. Blair, S. Gras, S.P. Vyatchanin,  Physics Letters A372,  5727, 2008 ;

%\bibitem{cree} Chree C., Quart. J. Pure and Appl.
%Math.,21, 83/84, 287-298 (1886).

% \bibitem{lamb} Lamb H., Proc. Roy. Soc. Lond. A93, 648, 114-121 (1917).


\bibitem{strigin} S.E.Strigin,  Physics Letters A372,  6305-6308, 2008 ;

\bibitem{thorne}    E. D'Ambrosio,  R. O'Shaughnessy, S. E. Strigin, K. Thorne and S. P. Vyatchanin,  http://arXiv.org: gr-qc/0409075;

\bibitem{thorne1}   R. O'Shaughnessy, S. E. Strigin and S. P. Vyatchanin, http://arXiv.org: gr-qc/0409050;

\bibitem{bond} M. Bondarescu and K.S. Thorne, Physical Review D74, 082003, 2006;

\bibitem{bond1} M. Bondarescu, O. Kogan, and Y. Chen, Physical Review D78, 082002, 2008;

\bibitem{mours} B. Mours, E. Tournefier, and J.-Y. Vinet, Classical and Quantum Gravity 23, 5777, 2006;

\bibitem{arlt}J. Arlt, K. Dholakia, L. Allen, and M.J. Padgett, Journal of Modern Optics 45, 1231, 1998;
\bibitem{clif}M. A. Clifford, J. Arlt, J. Courtial, and K. Dholakia, Optics Communications 156, 300,1998;
\bibitem{kenn} S.A. Kennedy, M. J. Szabo, H. Teslow, J.Z. Porterfield, and E.R. Abraham, Phys. Rev. A 66, 043801, 2002;
\bibitem{cour} J. Courtial and M.J. Padgett, Optics Communications 159,13 ,1999;
\bibitem{neil} A.T. O'Neil and J. Courtial, Optics Communications 181, 35,2000;
\bibitem{chu} S.-C. Chu and K. Otsuka, Optics Communications 281, 1647,2008;

\bibitem{hild} S.Chelkowski, S. Hild, A. Freise, arXiv:0901.4931v1[gr-qs], 2009.
\bibitem{abbott}B.P. Abbott et al., Reports on Progress in Physics 72, 076901, 2009;
\bibitem{levin}  V.B. Braginsky, Yu. Levin, S.P. Vyatchanin, Meas. Sci. Technol. 10 (1999) 598.
\bibitem{gbz} S. Gras, D.G. Blair, and C. Zhao, Classical and Quantum Gravity 26, 135012, 2009;
\bibitem{GSV} A.G. Gurkovsky, S.E. Strigin and S. P. Vyatchanin,
Physics Letters A  {\bf 362}, 91 (2007);

\bibitem{sv}
 S.E.~Strigin and S.P.~Vyatchanin, Physics Letters A {\bf 365}, 10 (2007).

\bibitem{sv1}  S.E.~Strigin and S.P.~Vyatchanin, Quantum
Electronics, {\bf 37}, No.12, 1097, (2007).


\bibitem{cqg} S. Gras, C. Zhao,  D.G. Blair, and L. Ju, Classical and Quantum Gravity, in press.
\bibitem{frit} M. Evans, L. Barsotti, and P. Fritschel, Physics Letters A, in press(doi: 10.1016/j.physleta.2009.11.023);
\end{thebibliography}
\end{document}